\documentstyle[11pt,aaspp4]{article}

\def\la{\mathrel{\hbox{\rlap{\hbox{\lower4pt\hbox{$\sim$}}}\hbox{$<$}}}}
\def\ga{\mathrel{\hbox{\rlap{\hbox{\lower4pt\hbox{$\sim$}}}\hbox{$>$}}}}

\slugcomment{Accepted to {\it The Astrophysical Journal (Letters)}}

\begin{document}

\title{The Redshift of GRB 970508}

\author{Daniel E. Reichart}

\affil{Department of Astronomy and Astrophysics, University of Chicago, Chicago, IL 60637} 

\begin{abstract}

GRB 970508 is the second gamma-ray burst (GRB) for which an optical afterglow has been detected.  It is the first GRB for which a distance scale has been determined:  absorption and emission features in spectra of the optical afterglow place GRB 970508 at a redshift of $z \ge 0.835$ (Metzger et al. 1997a, 1997b).  The lack of a Lyman-$\alpha$ forest in these spectra further constrains this redshift to be less than $z \sim 2.3$.  I show that the spectrum of the optical afterglow of GRB 970508, once corrected for Galactic absorption, is inconsistent with the relativistic blast-wave model unless a second, redshifted source of extinction is introduced.  
This second source of extinction may be the yet unobserved host galaxy.  I determine its redshift to be $z = 1.09^{+0.14}_{-0.41}$, which is consistent with the observed redshift of $z = 0.835$.  Redshifts greater than $z = 1.40$ are ruled out at the 3 $\sigma$ confidence level.

\end{abstract}


\section{Introduction}

Discovered by the BeppoSAX Gamma-Ray Burst Monitor (Costa et al. 1997), GRB 970508 is the second gamma-ray burst (GRB) for which an optical afterglow has been detected (e.g., Bond 1997).  Transient X-ray, near-infrared, millimeter, and radio emission have also been detected.  
GRB 970508 is the first GRB for which a distance scale has been determined:  Metzger et al. (1997a, 1997b) report the existence of absorption and emission features in spectra of the optical afterglow taken with the Keck II 10-m telescope $\approx$ 2 days and $\approx$ 26 days after the GRB event.  The identification of these features places GRB 970508 at a redshift of $z \ge 0.835$.  The lack of a Lyman-$\alpha$ forest in these spectra further constrains this redshift to be less than $z \sim$ 2.3.  Consequently, GRB 970508 is almost certainly cosmological in origin.

A host galaxy for GRB 970508 has not yet been observed.  However, Metzger et al. (1997a) report that the line emission observed in the Keck II spectra is consistent with constancy between their May 11 and June 5 observations.  Over this same period, the continuum emission has faded.  This suggests that a host galaxy of relatively weak continuum emission may be present at the observed redshift of $z = 0.835$.  (Metzger et al. 1997b).  

GRB afterglows are believed to be described by the relativistic blast-wave model (Paczy\'nski \& Rhoads 1993, Katz 1994a, Katz 1994b, M\'esz\'aros \& Rees 1997, Sari \& Piran 1997, Vietri 1997, Waxman 1997, Sari 1997, Katz \& Piran 1997).  Furthermore, the afterglow of GRB 970228, the only other GRB for which sufficient afterglow information is available, has been shown to be compatible with this model (Katz, Piran, \& Sari 1997, Waxman 1997, Wijers, Rees, \& M\'esz\'aros 1997, Reichart 1997, Sahu et al. 1997, Katz \& Piran 1997).  
A basic prediction of the relativistic blast-wave model is that after an initial period of increasing optical flux, lasting hours to days, the optical flux of the afterglow will decrease as a power-law of index $-1.5 \la b \la -0.5$.  
During this period of declining optical flux, the optical spectrum will be described by a power-law of index $a = 2b/3$ (e.g., Wijers, Rees, \& M\'esz\'aros 1997).\footnote{Other relationships between $a$ and $b$ can be found in the literature, however, these relationships refer to the GRB itself and not to its afterglow.  For example, internal collisions in a relativistic blast-wave could result in a premature GRB whose flux would decline much more rapidly than that given by $b = 3a/2$ (e.g., M\'esz\'aros \& Rees 1997); however, the persistent afterglow is produced when the blast-wave collides with the interstellar medium, which occurs whether or not the GRB is premature, and its decline in flux is given by this relationship (e.g., Waxman 1997).}
However, in the case of GRB 970508, $a \approx b$ (Sokolov et al. 1997), which is inconsistent with the relativistic blast-wave model. 

In \S2, I show that the reported GRB 970508 optical afterglow measurements, once corrected for Galactic absorption, are inconsistent with this prediction of the relativistic blast-wave model unless a second, redshifted source of extinction is introduced.  
This second source of extinction may be the yet unobserved host galaxy.  I determine its redshift and I estimate its hydrogen column density along the line of sight.  An observing strategy for future optical afterglow observations is recommended in \S3.
 
\section{Data Analysis \& Model Fit}

As of 1997 October 13, more than seventy optical and near-infrared measurements of the GRB 970508 afterglow have been reported.  These measurements span $>$ 80 days, the earliest of which was taken $\approx$ 4 hours after the GRB event.  
Although the photometry is generally quite good, with quoted errors that are often less than 0.1 mag, zero-point errors are evident between different observing groups in various bands.  Consequently, I consider only the largest self-consistent subset of these data:  Sokolov et al. (1997) report 22 optical measurements (B, V, R$_C$, and I$_C$ bands) taken with a 6-m telescope between $\approx$ 2 days and $\approx$ 31 days after the GRB event.  All of these measurements were taken after the optical flux had peaked (\S1), also $\approx$ 2 days after the GRB event.  These measurements are listed in Table 1.

I have corrected these measurements for Galactic absorption using the IRAS 100 $\mu$m V-band absorption measure of Rowan-Robinson et al. (1991) and the interstellar absorption curve (Johnson 1965, Bless \& Savage 1972).  
The four relevant corrections are $A_B = 0.12$, $A_V = 0.09$, $A_{R_C} = 0.06$, and $A_{I_C} = 0.04$ mag.  
I use the absorption measure of Rowan-Robinson et al., which measures the dust directly, instead of measures of the hydrogen column density, i.e. Hartmann \& Burton (1995), Stark et al. (1992), and Burstein \& Heiles (1982), since the IRAS 100 $\mu$m flux about the location of GRB 970508 varies significantly on angular scales that are smaller than the scales over which these measures of the hydrogen column density are averaged.  The corrected measurements are plotted in Figure 1.

The following model is now $\chi^2$-fitted to these corrected measurements:
\begin{equation}
F_\nu = F_0\nu^{\frac{2b}{3}}t^b - F_{ext}(\nu;A_V(z),z).
\end{equation}
The first term is the extinction-free prediction of the relativistic blast-wave model (\S1).  The second term is the correction that a second, redshifted source of extinction introduces.  It is given by redshifting the interstellar absorption curve and by specifying the magnitude of the extinction, which I parameterize as the V-band absorption magnitude {\it at the redshift of the source}:  $A_V(z)$.  A constant error of $\approx$ 0.07 mag must be added in quadrature to the quoted errors of Table 1 for the model to fit the data ($\chi^2 \approx \nu$).  
This suggests that either the quoted errors are underestimated by a factor $\sim 2$, or that the flux is varying by $\approx$ 6\% on timescales of days.
Reichart (1997) noticed possibly related temporal behavior in the light-curve of GRB 970228.  

The best fit is:  $\log F_0 = -9.03^{+0.44}_{-0.44}$ cgs, $b = -1.22^{+0.03}_{-0.03}$, $A_V(z) = 0.24^{+0.12}_{-0.08}$ mag, and $z = 1.09^{+0.14}_{-0.41}$.  
The quoted uncertainties are 1 $\sigma$ confidence intervals for one interesting parameter.  The best-fit redshift is consistent with the observed redshift of $z = 0.835$.  Furthermore, redshifts greater than $z = 1.40$ are ruled out at the 3 $\sigma$ confidence level.  The best-fit V-band absorption magnitude corresponds to a hydrogen column density of $\approx$ 4.5 x 10$^{20}$ cm$^{-2}$.  The possibility that there is no second source of absorption, i.e., $A_V(z) = 0$, is ruled out at the 3.8 $\sigma$ confidence level.

\section{Discussion \& Conclusions}

Using Equation (1) and the best-fit temporal decline power-law index, $b = -1.22$, I have scaled each of the measurements fitted to in \S2 to its corresponding value for May 11, just shortly after the optical peak.  These points define the post-peak, time-independent optical spectrum and are plotted in Figure 2.  The best-fit spectrum, as well as the extinction-free, relativistic blast-wave component of this spectrum (the first term of Equation (1)), are also plotted in Figure 2.  
A broad absorption feature is apparent in the best-fit spectrum.  This is the ultraviolet absorption feature of the interstellar absorption curve, redshifted into the B band.  Had the redshift of GRB 970508 been less than $z = 0.835$, U-band measurements would also have been necessary for this, the only strong feature of the interstellar absorption curve, to have been detected.
Consequently, in this type of analysis, different bands most sensitively probe different redshift ranges, depending on whether or not this absorption feature has been redshifted into the band in question.  Table 2 lists the redshifts that the five standard bands probe most sensitively.
Since cosmological GRBs are generally believed to have redshifts of $z \sim 1$, self-consistent sets of U-, B-, and V-band measurements should be a goal of future optical afterglow observations.

In this letter, I present a method by which redshifts can be determined for cosmological GRBs that are associated with host galaxies, even if absorption or emission lines are not observable or if spectra of sufficient quality are unattainable.  This method also yields hydrogen column densities along the line of site, which provides valuable information about the distribution of cosmological GRBs within galaxies.  For GRB 970508, I find that the redshift of its host galaxy, or possibly that of an intermediate galaxy, is $z = 1.09^{+0.14}_{-0.41}$, which is consistent with the observed redshift of Metzger et al.:  $z = 0.835$.  Redshifts greater than $z = 1.40$ are ruled out at the 3 $\sigma$ confidence level.  

\acknowledgments
This research has been supported by NASA grant NAG5-2868.  
I am also grateful to D. Q. Lamb, R. C. Nichol, B. P. Holden, F. J. Castander, W. B. Burton, and D. Hartmann for valuable discussions and information.

\clearpage

\begin{deluxetable}{ccc}
\tablecolumns{3}
\tablewidth{0pc}
\tablecaption{Optical Observations of the GRB 970508 Afterglow\tablenotemark{a}}
\tablehead{
\colhead{Band} & \colhead{Date\tablenotemark{b}} & \colhead{Magnitude\tablenotemark{c}}} 
\startdata
B & May 10.77 & $20.50 \pm 0.03$ \nl
B & May 10.93 & $20.60 \pm 0.03$ \nl
B & May 11.76 & $21.03 \pm 0.04$ \nl
B & May 12.87 & $21.48 \pm 0.06$ \nl
B & May 13.88 & $21.92 \pm 0.07$ \nl
V & May 10.77 & $20.06 \pm 0.03$ \nl
V & May 10.93 & $20.22 \pm 0.03$ \nl
V & May 11.76 & $20.52 \pm 0.03$ \nl
V & May 12.87 & $21.10 \pm 0.04$ \nl
V & May 13.88 & $21.47 \pm 0.05$ \nl
R$_C$ & May 10.77 & $19.70 \pm 0.03$ \nl
R$_C$ & May 10.93 & $19.80 \pm 0.03$ \nl
R$_C$ & May 11.76 & $20.10 \pm 0.03$ \nl
R$_C$ & May 12.87 & $20.63 \pm 0.05$ \nl
R$_C$ & May 13.88 & $21.09 \pm 0.07$ \nl
R$_C$ & May 22.00 & $22.20 \pm 0.15$ \nl
R$_C$ & Jun 9.60 & $23.28 \pm 0.10$ \nl
I$_C$ & May 10.77 & $19.19 \pm 0.04$ \nl
I$_C$ & May 10.93 & $19.30 \pm 0.03$ \nl
I$_C$ & May 11.76 & $19.58 \pm 0.04$ \nl
I$_C$ & May 12.87 & $20.19 \pm 0.06$ \nl
I$_C$ & May 13.88 & $20.58 \pm 0.09$ \nl
\enddata
\tablenotetext{a}{Sokolov et al. (1997), 6-m telescope.}
\tablenotetext{b}{1997 May 10.77 - June 9.60, UT in decimal days.}
\tablenotetext{c}{Uncorrected for Galactic absorption.}
\end{deluxetable}

\clearpage

\begin{deluxetable}{cc}
\tablecolumns{4}
\tablewidth{0pc}
\tablecaption{Redshifts That Standard Bands Probe Most Sensitively}
\tablehead{
\colhead{Band} & \colhead{$z$}} 
\startdata
U & 0.68 \nl
B & 1.05 \nl
V & 1.53 \nl
R$_C$ & 2.03 \nl
I$_C$ & 2.71 \nl 
\enddata
\end{deluxetable}

\clearpage



\clearpage

\figcaption[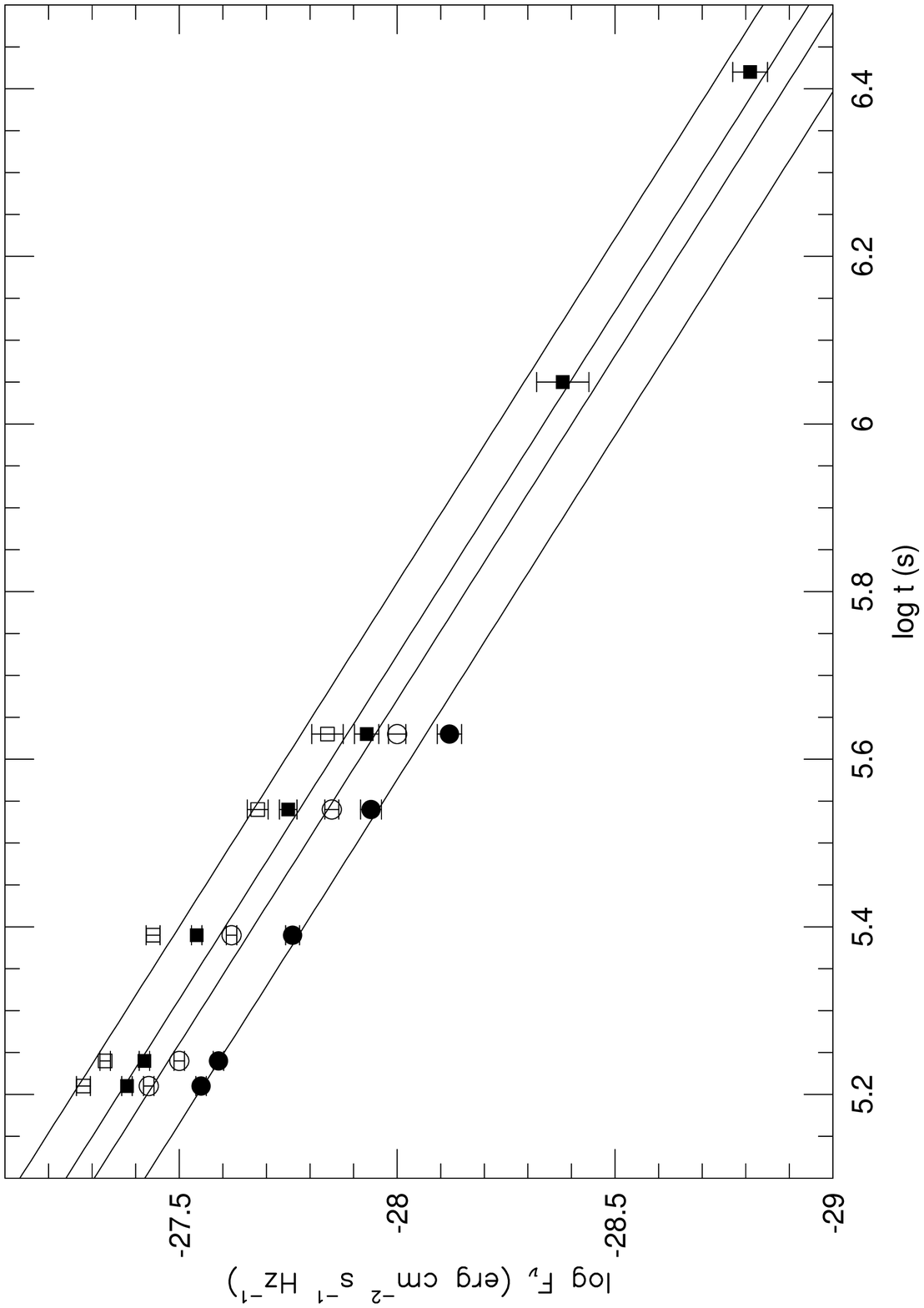]{Fluxes of Table 1 corrected for Galactic absorption (\S2) and the best fit to equation (1).  Solid circles are B band, open circles are V band, solid squares are R$_C$ band, and open squares are I$_C$ band.\label{fig1.ps}}

\figcaption[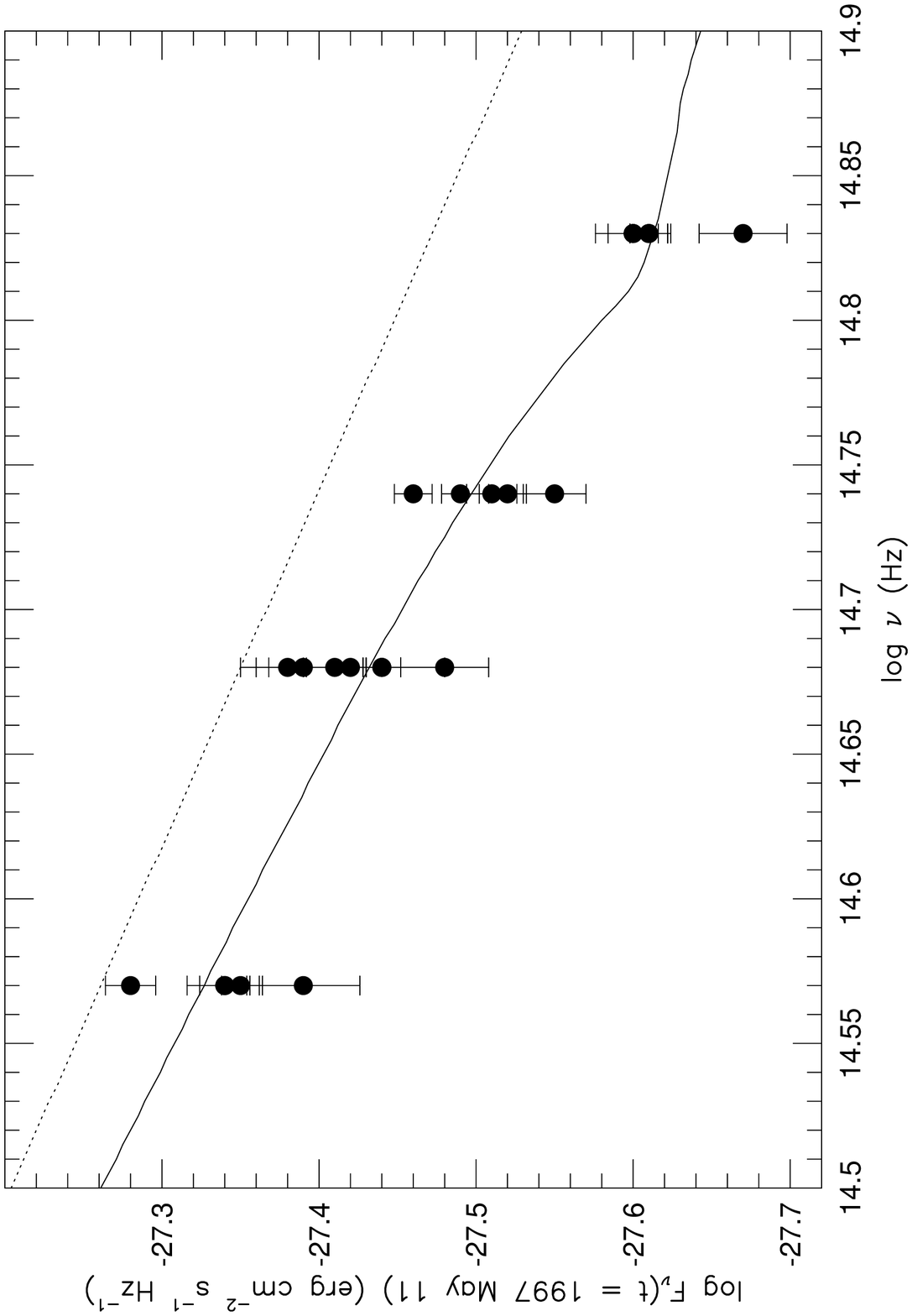]{The post-peak, time-independent optical spectrum of the afterglow of GRB 970508, scaled to 1997 May 11.  Plotted from left to right are the I$_C$-, R$_C$-, V-, and B-band fluxes of Table 1 corrected for Galactic absorption.  The solid line is the best-fit spectrum and the dotted line is the extinction-free, relativistic blast-wave component of this spectrum.\label{fig2.ps}}

\clearpage 

\setcounter{figure}{0}
\begin{figure}[tb]
\plotone{fig1.ps}
\end{figure}

\begin{figure}[tb]
\plotone{fig2.ps}
\end{figure}

\end{document}